\documentclass[twocolumn,aps,footinbib,pra]{revtex4}
\usepackage{amssymb}
\usepackage{amssymb}
\usepackage{graphicx}
\usepackage{amsmath}
\usepackage{epsfig}
\usepackage{times}
\usepackage{color}
\usepackage{subfigure}
\usepackage{setspace}
\usepackage{dcolumn}
\usepackage{bm}

\begin{document}

\title{ Many-body protected entanglement generation in interacting spin systems}
\author{A. M. Rey $^{1}$, L. Jiang $^{2}$,  M. Fleischhauer $^{3}$, E. Demler $^{2}$  and M.D.
Lukin $^{1,2}$}
 \affiliation{$^{1}$ Institute for Theoretical
Atomic, Molecular and Optical Physics, Cambridge, MA, 02138.}
\affiliation{$^{2}$ Physics Department, Harvard University,
Cambridge, Massachusetts 02138, USA}
 \affiliation{$^{3}$Fachbereich Physik, Technische Universitat Kaiserslautern, D-67663
  Kaiserslautern, Germany}
\date{\today }

\begin{abstract}
We discuss a method to achieve  decoherence resistent entanglement
generation in two level  spin systems governed by  gapped and
multi-degenerate Hamiltonians. In such systems,  while the large
number of degrees of freedom in the ground state levels  allows to
create various quantum  superpositions,   the energy
 gap prevents  decoherence. We apply the protected evolution  to   achieve decoherence resistent
generation of many particle GHZ states and show it can
significantly increase the sensitivity in frequency spectroscopy. We
discuss how to engineer the desired   many-body protected manifold
in two specific physical systems, trapped ions and neutral atoms in
optical lattices, and present simple expressions for the fidelity of
GHZ generation under non-ideal conditions.
\end{abstract}
 \maketitle

\section{Introduction}

It is well-known that entangled atomic states (e.g. so-called spin
squeezed states) potentially allow to significantly improve
resolution in Ramsey spectroscopy \cite{Leibfried,Leibfried2}.
Entangled states are also a fundamental resource in quantum
information and quantum computation science \cite{Nielsen,Preskill}.
However, in practice entangled states are difficult to prepare  and
maintain as noise and decoherence rapidly collapses them into
classical statistical mixtures. Thus  one of the most important
challenges in modern quantum physics is the design of robust and
most-importantly decoherence resistant methods for entanglement
generation.

 We have recently    proposed a method  \cite{Rey} that allows
 for  noise resistent generation of entangled states. The method  uses the
 energy gap of properly designed gapped-multi-degenerate  Hamiltonians.
 While the large number of degrees of freedom in the
 ground state manifold of such systems   allows to create various quantum
 superpositions  and to exploit rich dynamical evolution (suitable  for example  for precision spectroscopy), the energy
 gap prevents  decoherence as local excitations become energetically
 suppressed. A simple example of a many-body spin Hamiltonian that illustrates the idea of our scheme is
a multi-spin system with  isotropic ferromagnetic interactions.
These interactions will naturally align the spins. While all of the
spins can be rotated together around an arbitrary axis without  cost
of energy, local spin flips are energetically forbidden.

 This paper presents a detailed analysis of this method when applied
 to trapped ions and cold atoms in optical lattices.
We demonstrate
 its applicability for decoherence resistant
generation  of $N$-particle Greenberger-Horne-Zeilinger (GHZ) states
\cite{Greenberger} and  the potential of the latter to be used for
Heisenberg-limited spectroscopy. The paper is organized as follows:
In Sec II we review one of the standard procedures used to generate
multi-particle GHZ entangled states in ion traps and demonstrate the
detrimental effect of phase decoherence in such entanglement
generation schemes. In Sec III we explain the idea of a many-body
protected manifold (MPM) and discuss how and under what conditions
it
 can significatively reduce the effect of decoherence. In Sec. IV we show the applicability of the
gap protected evolution to precision measurements and discuss the
significant improvement in  phase sensitivity that it might provide.
In section V we elaborate on the physical resources required for the
implementation of the gap protected Hamiltonian in trapped ions and
discuss the advantages and disadvantages of its implementation with
respect to standard unprotected  Hamiltonians. In Sec. VI we study
how to engineer the long range interactions required for the gap
protected evolution in optical lattice systems interacting via short
range interactions and discuss the effectiveness of the MPM for GHZ
generation. In Sec. VII  we analyze in such systems the effect of
non-ideal conditions such as  the magnetic trapping confinement and
finally conclude in Sec.VIII.


\section{Multi-particle entanglement generation}


\subsection{Ideal Case}
In this section we start by reviewing a  method to generate
multi-particle entangled states  in a system  of $N$ spin 1/2 atoms
by time evolution  under  the so called squeezing Hamiltonian.
\begin{equation}
\hat{H}_z=\chi \hat{J}^{(0)2}_{z}. \label{hz}
\end{equation} As shown in Ref. \cite{Molmer,Sorensen} the $\hat{H}_z$ Hamiltonian can be
implemented in trapped ions  by using the collective
vibrational motion of the ions in a linear trap driven by
illuminating them with a laser field
 \cite{Leibfried, Leibfried2,Milburn}. In Eq.(\ref{hz}) we used
${\hat{J}^{(0)}} _{\alpha }$ to denote the collective spin operators
of the $N$ atoms: ${\hat{J}^{(0)}} _{\alpha }=\frac{1}{2}\sum
_i\hat{\sigma}_{i}^{\alpha }$, where $\alpha =x,y,z$ and
$\hat{\sigma}_{i}^{\alpha }$ is a   Pauli
 operator acting on the   $i^{th}$ atom and 
 we have identified the two relevant internal states of the
atoms with the effective spin index $\sigma ~=~\uparrow,\downarrow$.
In this manuscript we will use   units such that $\hbar =1$ and
assume $N$ to be even.

An appropriate basis to describe the dynamics of the system is the
one spanned by collective pseudo-spin  states   denoted as $
|J,M,\beta \rangle _{z}$ \cite{Arecchi72}. These states satisfy the
eigenvalue relations $ \hat{J}^{(0)2}|J,M,\beta \rangle
_{z}=J(J+1)|J,M,\beta \rangle _{z}$ and $
{\hat{J}^{(0)}}_{z}|J,M,\beta \rangle _{z}=M|J,M,\beta \rangle
_{z}$, with $ J=N/2,\dots ,0$ and $-J\leq M\leq J$.  $\beta $ is an
additional quantum number  associated with the permutation group
which  is required   to form a complete set of labels for  all the
$2^{N}$ possible states.

\begin{figure}[h] \addtolength{\belowcaptionskip}{-0.6
cm}\addtolength{\abovecaptionskip}{-0.6 cm}
\begin{center}
\leavevmode {\includegraphics[width=4in,height=2.8in]{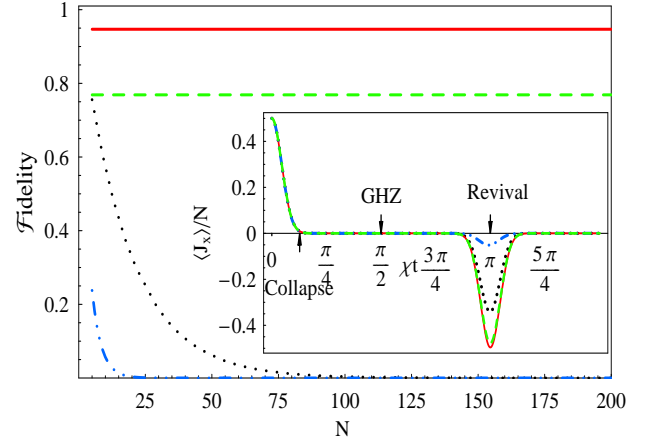}}
\end{center}
\caption{ (Color online) Fidelity of GHZ generation vs. $N$  with
and without gap protection. We assume a noise with step-like
spectral density with
 amplitude $f$ and cutoff frequency $\omega _{c}=\chi$ .
The dashed green and solid red lines are  for a protected system, $
\lambda=50$, with $f=0.6 \chi$
 and $f=0.1 \chi$ respectively. The dot-dashed blue and black dotted lines are for  an
unprotected system, $\lambda =0$, and  same decoherence parameters.
In the inset we show   $\langle {\hat{J}^{(0)}}_{x}(t) \rangle /N$
for $N=50 $. }\label{figu1}
\end{figure}

The entanglement  generation process starts by preparing the system
at $t=0$
 in a fully polarized state along the $x$ direction,
$|N/2,N/2\rangle _{x}$. As any  state with $J=N/2$ is uniquely
characterized by $M$, for denoting them we omit the additional
$\beta$ label. Fully polarized states along $x$
 can be written as a
superposition of states with different $M$ values along the $z$
direction: $\sum_M C_M|N/2,M\rangle _{z} $. During the evolution the
Hamiltonian
 imprints an $M^2$ dependent phase  to the different  components. As the system evolves,
  at  first the winding of the  phases  leads to a  collapse  of
  $\langle {\hat{J}^{(0)}}_{x}\rangle$. However, at time $\chi t_{rev}=\pi $ all  components  rephase  with
opposite polarization, and a perfect revival of the initial state
 is observed with   $\langle {\hat{J}^{(0)}}_{x}\rangle=-N/2$ (see
 Fig.\ref{figu1}). Specifically, the time evolution of  $\langle
 {\hat{J}^{(0)}}_{x}\rangle$ for systems with $N\gg1$ can be shown
 to be given by:

\begin{eqnarray}
\langle \hat{J}_{x}^{(0)}\rangle  &=&\frac{N}{2}%
\sum_{k=0,1,2,\cdots }(-1)^{k}e^{-N/2(\chi t-k\pi )^{2}} \label{jz2}
\end{eqnarray} Right at time  $t_0=t_{rev}/2$ the system becomes a macroscopic
superposition of fully polarized states along the $\pm x$ direction,
i.e. a $N$-particle GHZ state  of the form
\begin{equation}
|\psi ^{GHZ}_{x}\rangle \equiv \frac{1}{\sqrt{2}}\left( e^{-i\phi _{+}}|%
\frac{N}{2},\frac{N}{2}\rangle _{x}+e^{i\phi -}|\frac{N}{2},-\frac{N}{2}%
\rangle _{x}\right),\label{ghz}
\end{equation} with and $\phi _{\pm}$ real phases given by $-\pi/4$ and $\pi/4 +N
\pi/2$.

Recent experiments \cite{Leibfried, Leibfried2} have used this type
of scheme  to  generate GHZ states in trapped
 ions with the aim to perform precision measurements of $\omega_0$, the energy
 splitting between $\uparrow$ and $\downarrow$ levels. Ideally the use of GHZ states should  enhance the phase
sensitivity to the fundamental Heisenberg limit \cite{Bollinger}.
However, decoherence significantly limited the applicability of the
method.

\subsection{Effect of decoherence}

To understand the detrimental effect of decoherence   we first
assume that the dominant type of decoherence is
 single-particle dephasing. Such dephasing comes from  processes that, while preserving
the populations in the atomic levels, randomly change the phases
leading to  a decay of the off-diagonal density matrix elements. We
model the phase decoherence by adding to Eq. (1) the following
Hamiltonian \cite{Huelga}
\begin{equation}
\hat{H}_{env}=\frac{1}{2}\sum_{i}h_{i}(t)\hat{\sigma}_{i}^{z},\label{env}
\end{equation}
where $h_{i}(t)$ are assumed to be independent stochastic  Gaussian
processes with zero mean and with
 autocorrelation function $\overline{h_{i}(t)h_{j}(\tau )}%
=\delta _{ij}f(t-\tau )$. Here the bar denotes averaging over the
different random outcomes. In what follows we will use the property
that zero mean  Gaussian variables satisfy  $\overline{%
\exp [-i\int_{0}^{t}d\tau h(\tau )]}=\exp [-\Gamma(t)]$, with
$\Gamma(t)
=\frac{1}{2}\int_{0}^{t}dt_{1}\int_{0}^{t}dt_{2}f(t_{1}-t_{2})$
\cite{Papoulis}.

\begin{figure*}[tbh]
\begin{center}
\leavevmode {\includegraphics[width=6 in]{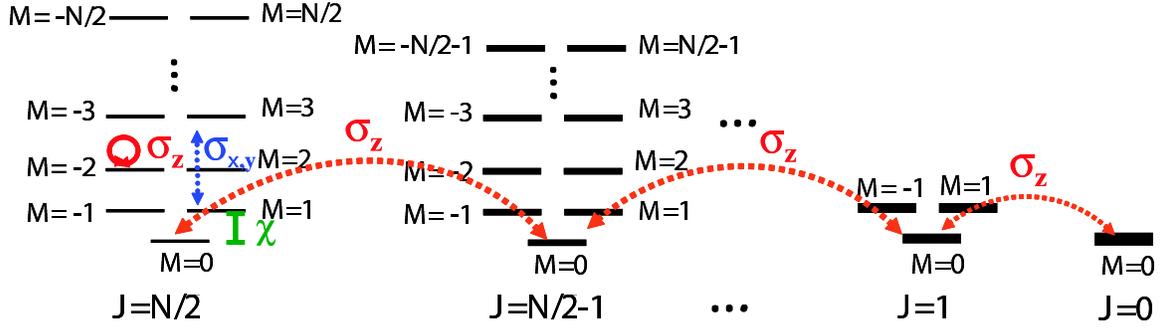}}
\end{center}
\caption{ Schematic representation of the energy levels of the $\chi
\hat{ J}_z^{(0)2}$ Hamiltonian and the effect of the different type
of noise. As states with different $J$ but equal $|M|$ are
degenerate,  in the presence of phase decoherence ($\sigma_z$ noise)
they are populated during the time evolution. $\sigma_{x,y}$ noises
couple states which  differ by $\pm 1$ units of M and therefore  the
small energy gap between them, of the order of $\chi$, naturally
protects the system from these type of processes.}\label{figu2}
\end{figure*}

Phase decoherence causes an  exponential  decay of the revival peak
and  and  $N$ dependent decay of the fidelity, defined as
$\mathcal{F}(t_0)=\overline{\langle \psi ^{GHZ}_x
|\hat{\rho}(t_0)|\psi ^{GHZ}_{x}\rangle }$ (See Fig. \ref{figu1}).
Qualitatively the effect of phase decoherence on the evolving state
can be understood from the energy levels of $\hat{H}_z$ (See Fig.
\ref{figu2}). While  $\hat{H}_z$  commutes  with both $
\hat{J}^{(0)}_z$ and $ \hat{J}^{(0)2}$, $\hat{H}_{env}$ only commutes
with $ \hat{J}^{(0)}_z$   does not commute with $
\hat{J}^{(0)2}$. Therefore, in the presence of phase decoherence
transitions between different $J$ subspaces are allowed as long as
$M$ is conserved. As all the states with the same $|M|$ value are
degenerate, there is no energy barrier to prevent such transitions
and very quickly the initially populated $J=N/2$ manifold is
depleted and the fidelity of generating the GHZ state significantly
degraded. Moreover, the degradation  scales  exponential with
increasing $N$ due to the exponential scalability  of the number of
accessible states to which the initially $J=N/2$ population can be
transferred to.

Quantitatively the effect of decoherence can be calculated by using
the uncoupled spin basis as it diagonalizes the total  Hamiltonian.
Each state in this basis can be label as $\{|n^{(k)}\rangle
=|s_{1}^{k},s_{2}^{k},\dots ,s_{N}^{k}\rangle \}$, where
$s_{i}^{k}=\pm 1$ for $\uparrow \downarrow $ and $k=1,\dots 2^{N}$.

If at $t=0$ the reduced density matrix of the system, $\hat{\rho}$,
is given by  $\hat{\rho}=\sum_{k,l}
\rho_{k,l}(0)|n^{(k)}\rangle\langle n^{(l)}|$,
 after time $t$
 \begin{eqnarray}
 \rho_{k,l}(t)&=&  \rho_{k,l}(0) e^{\frac{i}{2} \chi t [(\sum_{i=1}^N s_{i}^{k})^2- (\sum_{i=1}^N
 s_{i}^{l})^2]}\times\notag\\&&
e^{-i\sum_{i=1}^N \int_0^t d\tau h_i(\tau) (s _{i}^{k}- s_{i}^{l})}
\label{dede}
 \end{eqnarray}
 As a consequence, $\langle {\hat{J}^{(0)}}%
_{x}(t)\rangle = Tr[
\overline{{\hat{J}^{(0)}}_{x}\hat{\rho}(t)}]=e^{-\Gamma(t)}\langle
{\hat{J}^{(0)}}_{x}(t)\rangle|_{\Gamma =0}$ with $\langle
{\hat{J}^{(0)}}_{x}(t)\rangle|_{\Gamma =0}$ the expectation value in
the absence of noise (Eq.(\ref{jz2})). The factor $e^{-\Gamma(t)}$
comes from the fact that  the operator
${\hat{J}^{(0)}}_{x}=\sum_{i}\hat{\sigma}_{i}^{x}$ only probes
one-particle coherence, i.e it only  connects states with exactly
one spin flipped.

Assuming  that at $t=0$ all atoms are polarized in the $x$
direction, i.e. $\rho_{k,l}(0)=2^{-N}$, one can show from Eq. (\ref
{dede}) that the fidelity is degraded to

\begin{equation}
\mathcal{F}(t_o)=\frac{1}{2^{2N}}\sum_{l,k}
e^{-\frac{\Gamma(t_o)}{4}\sum_{i=1}^N (s _{i}^{k}- s_{i}^{l})^2}
=\left(\frac{1+ e^{-\Gamma(t_o)}}{2}\right)^N \label{dec}
 \end{equation}
 %
 %


\section{Protected dynamics}


\subsection{ Manybody protected manifold (MPM)}

Let us now consider what happens if in addition of $\hat{H}_z $ we
assume that there is an  isotropic infinite range ferromagnetic
interactions between the spins so the system Hamiltonian is
described by the Hamiltonian $\hat{H_c}=\hat{H}_{prot}+\hat{H}_z $,
where
\begin{eqnarray}
\hat{H}_{prot}&=&-\lambda \hat{J}^{(0)2} \label{Hs}
\end{eqnarray} Here we have assumed that all atoms are in different orbitals  but have enough
of spatial overlap   that every spin interacts with every other
spin.

The isotropic Hamiltonian $\hat{H}_{prot}$  has a  ground state
manifold spanned by a set of $N+1$ degenerate states.   They  lie on
the surface  of  the  Bloch sphere with maximal radius $J=N/2$
 and are  totally  symmetric, i.e. invariant with respect to particle
permutations.  There is a finite energy gap $E_g=\lambda N$ that
isolates the ground state manifold from the rest of the Hilbert
space. This gap is the key for the many-body protection  against
decoherence. Hereunder we will refer to the ground state manifold as
the \textit{many-body protected manifold} (MPM).

\subsection{Protection against phase decohernce}

\begin{figure}[tbh]
\begin{center}
\leavevmode {\includegraphics[width=3.5 in]{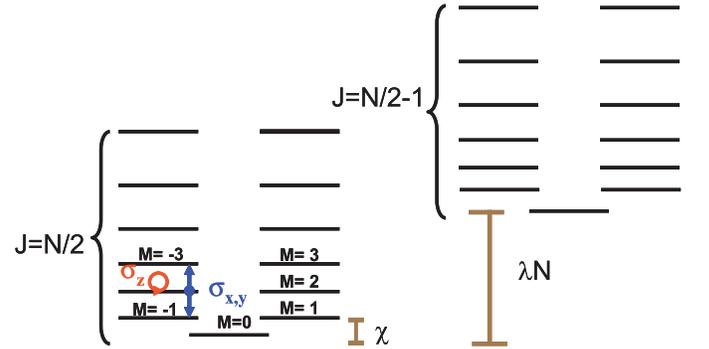}}
\end{center}
\caption{ Schematic representation of the energy levels of the
$\hat{H}_{prot}+ \chi \hat{ J}_z^{(0)2}$ Hamiltonian.
$\hat{H}_{prot}$ lifts the degeneracy of the different  $J$
manifolds and suppresses (in the slow noise limit) couplings between
them. So if a $t=0$ the system is in the MPM it remains there.
}\label{figu2b}
\end{figure}

The  low energy  spectrum of $\hat{H_c}$ is shown in Fig.
\ref{figu2b}. As $\hat{H}_{z}$ commutes with $\hat{H}_{prot}$, in
the absence of decoherence the latter does not affect at all the GHZ
generation dynamics, however in the presence of decoherence the
latter does significatively reduces the effect of local
environmental noise. The protection  can be best  understood by
using the basis of collective states. In terms of collective spin
operators $\hat{H}_{env}$ can be written as:

\begin{equation}
\hat{H}_{env}=\frac{1}{\sqrt{N}}\sum_{k=0}^{N-1}g^{k}(t)\hat{J}^{(k)}_z,\label{env2}
\end{equation}
where $g^{k}(t)=\frac{1}{\sqrt{N}}\sum h_j(t) e^{-i\frac{2 \pi j k}{N}}$ and
 $\hat{J}^{(k)}_{\alpha}=\frac{1}{2} \sum \hat{\sigma}_j e^{i\frac{2 \pi j k}{N}}$.
Note that allowed transitions must conserve $M$ as both the system
and noise Hamiltonian commute with $\hat{J}_{z}^{(0)}$. In the
presence of a large energy gap $E_g$, one can distinguish two
different type of  processes: (i) Decoherence effects that take
place  within the MPM
 due to the collective dynamics induced by  the $k=0$ component of  $\hat{H}_{env}$,
  and (ii) transitions   across the gap induced by the inhomogeneous terms.
The later couple  the MPM with the
 rest of the system, however they are nonenergy conserving process
and consequently   perturbatively weak.

 Using a perturbative analysis, and assuming that at $t=0$ the
system lies within the MPM, the evolution  of the projection of the
density matrix
 on the MPM: $ \rho_{M \tilde{M}}\equiv {}_z\langle
N/2,\tilde{M}|\hat{\rho}| N/2,M \rangle_z$ can be written as
\begin{equation}
 \rho_{M,\tilde{M}}(t)= \rho_{M,\tilde{M}}(0) e^{i t \chi (M^2-\tilde{M}^2)}
 e^{i(\theta_M-\theta_{\tilde{M}}}) e^{-\frac{1}{2}(\gamma_M-\gamma_{\tilde{M}}}) ,\label{densp}
\end{equation}Here

\begin{equation}
\theta _{M}(t)\equiv \langle \frac{N}{2},M|\int_{0}^{t}d\tau \hat{H}%
_{env}(\tau )|\frac{N}{2},M\rangle
=\frac{M}{\sqrt{N}}\int_{0}^{\tau}g^0 (\tau),
\label{thet}\end{equation} accounts for the dynamics induced by the
noise within the MPM and
\begin{equation}\gamma^{M}(t)=\sum_{J\neq{N/2},\beta }|\int_{0}^{t}d\tau
\mathcal{M}_{J,\beta}^{M}e^{i\tau
\omega_{J,\beta}}|^{2},\end{equation} takes into account the
depletion of the $J=N/2$ levels due to transition matrix elements
between $|\frac{N}{2},M \rangle_z $  and states outside the MPM:
 $\mathcal{M}_{J,\beta }^{M}={}_z\langle
\frac{N}{2},M|\hat{H}_{env}|J,M,\beta \rangle {}_z$.
$\omega_{J,\beta}$ are the respective energy splittings. Because
$\hat{H}_{env}$ is a vector operator, according to the Wigner-Eckart
theorem, $\hat{H}_{env}$ only couples the states in the MPM  with
states which have  $J=N/2-1$
 and thus with excitation energy $\lambda N$.

 Assuming the power spectrum of
the noise, $f(\omega )\equiv \int dte^{-i\omega t}f(t)$, to have a
cut-off frequency $\omega _{c}$ (e.g. $f(\omega )= f $ for
$\omega\leq \omega_c$ and 0 otherwise),
 we find that
\begin{equation}
\overline{ \gamma
^{M}(t)}\approx \frac{N^{2}-4M^{2}}{N}f \int_{0}^{\omega _{c}}d\omega \left(%
\frac{\sin (t(\omega -\lambda N)/2)}{\omega -\lambda N}\right)^{2}
.\label{gampr}
\end{equation}

In the limit when the noise is sufficiently slow , i.e. $\omega_c
\ll E_g$, then $\overline{ \gamma ^{M}(t)}$ is bounded for all
times, $\overline{ \gamma
^{M}(t)}<(\frac{N^{2}-4M^{2}}{N^{2}})(\frac{f \omega _{c}}{\lambda ^{2}N}%
)\ll 1 $ and the atomic population within the ground state manifold
is fully preserved i.e.
 $\gamma ^{M}(t)\approx0$ in Eq. (\ref{densp}).

Consequently, in the slow noise limit type (ii) processes are
energetically forbidden and
  only type (i) processes are effective and therefore the noise acts  just as
a uniform  random magnetic field:  if at $t=0$
$\hat{\rho}=\sum_{M,\tilde{M}}
\rho_{M,\tilde{M}}(0)|\frac{N}{2},\tilde{M}\rangle\langle\frac{N}{2},M|$,
then after time $t$ each component $\rho_{M,\tilde{M}}$  acquires an
additional random  phase $e^{i (\theta _{M}(t)-\theta
_{\tilde{M}}(t))}$ and on average
\begin{eqnarray}
\overline{ \rho_{M,\tilde{M}}(t)}&=& \rho_{M,\tilde{M}}(0) e^{i \chi
t(M^2-\tilde{M})^2} e^{-\Gamma(t)\frac{(M-\tilde{M})^2}{N}}.
\label{pde}
\end{eqnarray}
 The factor of $\sqrt{N}$ in the denominator of $\theta_M$ is  fundamental for the
reduction of the effect of decoherence within the MPM. For example,
it  makes ${\hat{J}^{(0)}}_{x,y}$  to decay $N$ times slower than in
the unprotected system: i.e.
$\langle{\hat{J}^{(0)}}_{x,y}(t)\rangle=e^{-\Gamma(t)/N}
 \langle{\hat{J}^{(0)}}_{x,y}(t)\rangle|_{\Gamma=0}$.

Assuming  all atoms  are initially polarized in the $x$ direction,
$\rho_{M,\tilde{M}}(0)=2^{-N} \sqrt{\binom{N}{M+N/2}
\binom{N}{\tilde{M}+N/2}}$, using Eqs. (\ref{pde}), the
approximation
  $\binom{N}{M+N/2}\approx  (\frac{2}{\pi N })^{1/4}e^{-\frac{ M^2}{N}}$, valid in the large $N$ limit
  and replacing  the sums over $M$ and $\tilde{M}$  by integrals
the fidelity at a given time $t$ can be shown to be  given by:

\begin{equation}
\mathcal{F}(t_o)=\frac{1}{\sqrt{1+\Gamma(t_o)}}\label{fid}.
 \end{equation}
The insensitivity  of $\mathcal{F}(t)$  on  $N$, and the  $N$ times slower decay rate of  $\langle {\hat{J}^{(0)}}%
_{x}\rangle $ demonstrate  the  usefulness of MPM
 to generate  a large number of entangled particles.

\subsection{Protection against arbitrary noise}
We now  discuss the protection against spin flips, which can be
modeled by terms proportional to $\sigma_i^x$,
 $\sigma_i^y$ in the noise Hamiltonian, Eq.(\ref{env}). First of all
 note that as the $\uparrow$ and $\downarrow$
   states have a finite energy splitting   $\omega_0$,   low frequency noise
 associated with such terms will be suppressed  due it.
However, most of the spin flips are generally induced by
imperfections in the laser fields and therefore  are at frequencies
close to $\omega_0$, i.e. they correspond to  low frequency noise in
the rotating frame of the laser.  In the case involving GHZ state
generation,  the finite energy cost imposed by  $\chi J_z^2$ between
levels with different $|M|$ value tends to  inhibit these processes
as illustrated in Fig. 2. If in addition $\hat{H}_{prot} $ is
present  this natural protection can be enhanced due to the fact
that the energy gap suppresses  the component of the noise that
cause transitions between the MPM and other manifolds. More
precisely, noise modeled as $\sum_i h^\alpha \hat{\sigma}_i^\alpha$,
when projected into the MPM reduces to $\frac{1}{\sqrt{N}}
g^\alpha(0)\hat{J}^{0}_\alpha$ with $\alpha=x,y,z$.

\begin{figure}[tbh]
\begin{center}
\leavevmode {\includegraphics[width=3.5 in]{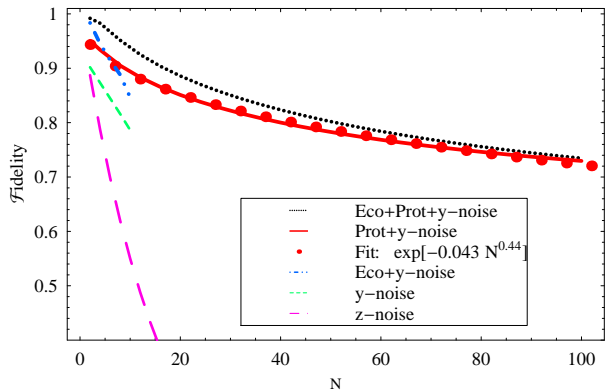}}
\end{center}
\caption{(Color Online) Fidelity to create GHZ state as a function
of N for systems in the presence of $\sigma_y$ noise : without MPM
(green dashed line), without MPM but with spin echo (dot-dashed blue
line), with MPM (red solid line) and with
 both MPM and echo (dotted black line). We also show the degradation
caused  by pure dephasing  in an unprotected system (long dashed
purple line) for comparison purposes. For simplicity we assumed
infinite correlation time: $f(\omega)=\Upsilon
\delta(\omega/\chi)$,$\Upsilon=0.1 \chi$. The echo technique
consisted of  a perfect sudden $\pi$ pulse around the $x$ direction
at $\chi t=\pi/4$. Because the  $y$ components of the noise do not
commute with  $\hat{J}_z^{2(0)}$ the dynamics was solved
numerically.  For the unprotected system all the $2^N$ states had to
be considered and  we  had to limit the particle number by
 $N=10$. On the other hand   the restriction of  the dynamics to
 the MPM in the protected scheme
 allowed us to extend the calculation to larger $N$ values.
 For the $\sigma_y$ noise  the
fidelity with protection does not become $N$ independent, but
instead  scales as $e^{-0.043 N^{0.44}}$ (see fitted red dots).
Nevertheless we do gain a factor of order $\sqrt{N}$ with respect to
the unprotected system. The plot also shows that
 the combination of   MPM with spin  echo  provides the best
protection. }\label{figur3}
\end{figure}

  In Fig. \ref{figur3} we quantify  the protection provided by the
  MPM. In the absence of any
protection we find the finite energy cost imposed by $H_{z}$ helps
to protect the system against transversal noise. Instead of the
exponential decay ( $\sim e^{-N \Gamma(t_0)}$ )  of the fidelity
observed when dephasing is present spin flips degrades the fidelity
as $\sim e^{-\Gamma(t_0)N/4}$  ( at least for the moderated $N<12$
we have to restrict our simulations). With protection the fidelity
scales even better as  the transversal noise is restricted to act
only within the MPM . Instead of the exponential decay  of the GHZ
generation fidelity with $N$, with protection it  decays as $\sim
e^{-A\Gamma(t_0) N^{0.44}}$ with $A$ a numerical constant, $A\sim
1/3$ (for this result we do not have to restrict to moderated N).
Hence,  with $\hat{H}_{prot}$ we gain a factor of order $\sim
\sqrt{N}$. The reason why $\hat{J}^{0}_x$ and $\hat{J}^{0}_y$ noise
degrade stronger the fidelity than $\hat{J}^{0}_z$ noise (which
leads just to a $N$ independent fidelity) is that   the former do
not commute with  $\hat{H}_z$  and mix states with different  $M$
quantum number. Additionally  in the figure we show  that
      for noise with long
correlation time the use of spin echo techniques  can help to
further reduce the effect of decoherence.

\section{Applications to precision measurements using trapped ions}

Recent experiments \cite{Leibfried, Leibfried2} have generated GHZ
states made of up to six beryllium ions and used them  to perform
precision measurements of $\omega_0$. For the ideal GHZ state
preparation, the  spectroscopy should lead to Heisenberg-limited
resolution, $|\delta \omega_0 |\propto N^{-1}$ \cite{Bollinger}.
However, in practice, even for six ions, the phase accuracy   was
significantly degraded by  decoherence.

The spectroscopy   \cite {Leibfried, Leibfried2} was realized  by
first  creating the desire GHZ state by applying to the initial
polarized state, $|J=N/2,N/2\rangle_z$ the unitary gate operation
  $U_N=e^{i \pi/2 {\hat{J}^{(0)}}_y} e^{i \pi/2  \hat{J}^{(0)2}_{z}}e^{-i \pi/2 {\hat{J}^{(0)}}_y}$.
Then   the GHZ state was let to  to freely  precess in the $z$
direction  for time
  $t$ so  each atom accumulated a phase difference $\phi=(\omega -\omega_0)t$ (in a reference   frame
rotating with the  frequency $\omega$, the frequency of  the applied
field).
 The phase difference
was then decoded  by measuring the collapse probability into the
states $|J=N/2,N/2\rangle_z$ or $|J=N/2,-N/2\rangle_z$ after
applying the unitary transformation $\prod_i\sigma^x_i$.

 This generalized
Ramsey sequence  can be quantitatively described  as a measure
of the expectation value of the following operator, $\hat{O}$:
\begin{equation}
\langle \hat{O}\rangle =\langle \psi (t_{0})|\prod_{i}[\cos (\phi )\hat{\sigma}_{i}^{z}-\sin (\phi )\hat{%
\sigma}_{i}^{y}]|\psi (t_{0})\rangle \label{eqO}
\end{equation}
with $|\psi (t_{0})\rangle ~=~e^{-i\frac{\pi }{2}
 \hat{J}^{(0)2}_{z}}|N/2,N/2\rangle _{x}.$

 The phase sensitivity   $\langle \Delta ^{2}\hat{O}\rangle$ achievable by
repeating the above scheme during total time $T$ is related to the
signal variance $\langle \Delta ^{2}\hat{O}\rangle =\langle
\hat{O}^{2}\rangle
-\langle \hat{O}\rangle ^{2}$ and given by: $|\delta \omega_0 |~=~\sqrt{\frac{1}{%
tT}\frac{\langle \Delta ^{2}\hat{O}\rangle }{(\delta \langle
\hat{O}\rangle
/\delta \phi )^{2}}}$ \cite{Huelga}. Because $%
\hat{O}^{2}=1$, we just have to calculate $\langle
\hat{O}\rangle={\rm Tr}[\overline{\hat{\rho}(t_0)\hat{O}}]$ to
evaluate $%
\delta \omega_0 $.

Experimentally, magnetic field noise  is one of the   sources of
phase decoherence. Assuming that such dephasing mainly takes  place
during the GHZ generation, as during the Ramsey interrogation  time
the atoms are essentially freely evolving, using   Eq.(\ref{dede})
one can show  that  for the
 unprotected system

\begin{eqnarray}
&&\langle \hat{O}\rangle =\langle \psi ^{GHZ}_x|\prod_{i}[\cos (\phi
)\hat{\sigma}_{i}^{z} -e^{-\Gamma(t_{0})} \sin (\phi
)\hat{\sigma}_{i}^{y}]|\psi ^{GHZ}_x\rangle \notag\\&&=
\frac{1}{2}[\frac{e^{i\phi }(1-e^{-\Gamma (t_{0})})+e^{-i\phi
}(1+e^{-\Gamma (t_{0})})}{2}]^N +h.c   \label{opt}
\end{eqnarray}
Consequently the maximal phase resolution, achieved at $\phi
_{opt}=n \pi $ (for integer $n$), can be shown to be given by
\begin{equation}
|\delta \omega_0 |_{opt}=|\delta \omega_0 |_{sh} /G,  \label{sens}
\end{equation} with $G=\sqrt{((N-1) e^{-2\Gamma(t_0)}+1)}$ and $ |\delta \omega_0
|_{sh}=\frac{1}{\sqrt{tTN}}$ the shot noise resolution. The factor
$G$ explains the strong limitations introduced by decoherence. If
the time required to generate the GHZ state is such that $G\sim 1 $
(i.e when $\Gamma(t_{0})>\ln [\sqrt{N}]$),
 the phase accuracy is reduced to the classical  shot noise
resolution.

 However, if instead $\hat{H}_{prot}+ \hat{H}_z$
 is used for the GHZ generation,  $G$  is replaced by
  $\sqrt{((N-1) e^{-2\Gamma(t_0)/N}+1)}$. Due to  the $N$ times slower  decay  rate of
the atomic coherences, the same preparation time that  leads to shot
noise resolution without protection, can lead to a  sensitivity  at
the level of the fundamental Heisenberg limit with protection.

\section{Implementation of the Gap protected Hamiltonian in trapped
ions}

We now proceed to  review and complement  the implementation of
the protected Hamiltonian, $\chi(\hat{J}^{(0)2}-\hat{J}_z^{(0)2} )$,
proposed in Ref.\cite{Unanyan}. Consider a linear trap with a string
of ions with two relevant internal levels. The ions are assumed to
be cooled such that only the in-phase collective center of mass
oscillation of all ions is excited. The corresponding oscillation
frequency is denoted by $\nu$. The two internal levels of the ions
are coupled by a laser field  with a slowly varying Rabi frequencies
$\Omega$ and with frequency $\omega_1=\omega_o-\delta$, being
$\delta$ the detunning from resonance. Assuming that the field
couple all ions in the same way, we can describe the system by the
Hamiltonian $\hat{H}=\hat{ H}_0 +\hat{H}_{in}$, where $H_0= \nu
\hat{a}^\dagger \hat{a} + \omega_o \hat{J}_z^{(0)}$, $\hat{a}$ being
the annihilation  operator of the quantized oscillation mode. The
interaction Hamiltonian $\hat{H}_{in}$ is given  by
\begin{equation}
\hat{H}_{in}=\Omega \hat{J}_+^{(0)}e^{i \delta t } e^{i \eta
(\hat{a}^\dagger e^{i \nu t } +\hat{a} e^{-i \nu t })}+h.c.
\label{int}
\end{equation}  where $\eta$ is the Lamb-Dicke parameter. The
 detuning  $\delta$ is assumed to be large compared to the
  linewidth of the resonance
but sufficiently different from the frequency of the center os mass
oscillation. As a consequence, the dominant processes are two-photon
transitions leading to a simultaneous excitation of pairs of ions.

We first assume  that the ion trap is in the Lamb-Dicke limit, i.e.,
that the ions are cooled sufficiently enough, such that for all
relevant excitation numbers $n$ of the trap oscillation $(n+
1)\eta^2 \ll1$ holds. In this limit one  can expand the exponent in
Eq. (\ref{int}) to first order in $\eta$. Confining the interest to
time averaged dynamics over a period much longer than any of the
oscillations present in $\hat{H}_{in}$, then the oscillatory terms
may be neglected and we are left with a more simple effective
Hamiltonian Ref.\cite{new} :

\begin{equation}
H_{eff}= \chi(\hat{J}^{(0)2}-\hat{J}_z^{(0)2} )+
\frac{2\Omega^2}{\delta} \hat{J}_z^{(0)} +\Lambda (2n+1)
\hat{J}_z^{(0)}\label{eff}
\end{equation}where $\chi=\frac{2\nu \eta^2 \Omega^2}{\delta^2-\nu^2}$, $\Lambda=\frac{\chi
\delta}{\nu}$ and $n$ the number of phonons in the vibrational mode.
The first term in $H_{eff}$ is the desire protected Hamiltonian. The
second term acts as an effective magnetic field which can be
canceled by adding an  external  magnetic field or by echo
techniques. The third term comes the  from  the ac Stark shift  of
the atomic levels due to the laser fields. In contrast to the
standard scheme used to create $\hat{J}_z^{(0)2}$, where the $n$
dependence exactly cancels, here it  does not and if not corrected
can certainly degrade  the fidelity. The degradation can be shown to
be given by:

\begin{equation}
\mathcal{F}(t_o)=\sum_n P_n\exp[-\frac{N^2\Lambda^2
(2n+1)^2t_o^2}{8}]
\end{equation}where $P_n$ is the initial population of the state with $n$ phonons.

In order to prevent this  effect one has to cool the ions to the
ground state, $P_n=\delta_{n0}$, which might be feasible with the
state of the art technology or alternatively one can  use spin echo
techniques. For example if at time $t_o/2$ the sign of the laser
detuning $\delta$ is changed, then the different components will
rotate in the opposite direction and at $t_o$ the net effect due to
the extra second and third terms in $\hat{H}_{eff}$ will be canceled
out.

So far we have used the Lamb-Dicke and the rotating-wave
approximation. Now we perform a more detailed analysis of the
validity of these approximations and estimate the effect of
deviations from the ideal situations in an actual experiment. To do
that  we follow Ref. \cite{Sorensen} and change to the interaction
picture of $\hat{H}_{eff}$, assuming that the undesired second and
third terms can be canceled by the techniques described above, and
treat the small non-ideal deviations by perturbation theory.
\begin{itemize}
  \item {\it{Direct coupling}}
  \end{itemize} Going from Eq.(\ref{int}) to Eq.(\ref{eff}) the off-resonant
term $H_d=\Omega \hat{J}_+^{(0)}e^{i  t \delta } +h.c.$ was
neglected. This term correspond to direct single atom spin flips
without any vibrational excitation.

Changing to the interaction picture of  $H_{eff} $ and  using the
fact that $H_d$ oscillates a much higher frequency that
$\hat{U}(t)=e^{i H_{eff} t}$ so that the latter can be treated as
constant in the integrals used  in  the Dyson series, one can show
that

\begin{equation}
\mathcal{F}(t_o)=1-\frac{\Omega^2}{\delta^2}\left(N^2\sin^2(\delta
t_o)+4 N\sin^4(\delta t_o/2)+\dots\right)
\end{equation}The degradation of fidelity is a factor of $N$ larger
than the degradation caused by direct couplings in the standard
realization of $J_z^{(0)2}$ where
$\mathcal{F}(t_o)=1-\frac{N\Omega^2}{\delta^2}\sin^2(\delta t_o)$.
Therefore it is important for the implementation of the protected
hamiltonian to use  weak laser power or to control the system
parameter such that $\delta t_0=2 K \pi$ with $K$ an integer.

\begin{itemize}
  \item {\it{Lamb-Dicke approximation}}
  \end{itemize}  In Ref.\cite{Sorensen} it has been shown that relaxing the Lamb-Dicke
approximation and including higher order terms results in an
effective  $\chi_n $ which depends on the vibrational number of
phonos in the collective  mode:
$\chi_n=\chi[1-\eta^2(2n+1)+\eta^4(5/4 n^2+5/4 n+1/2)]$. As this
effect is global, the gap does not protect against it and  it leads
to a degradation of the fidelity given by

\begin{eqnarray}
\mathcal{F}(t_o)&=&\sum_n P_n \left(1+\frac{N(N-1)(\pi/2-\chi_n
t_o)^2}{4} \right )^{-1/2}\notag\\&\sim&
1-\frac{\pi^2N(N-1)\eta^4}{32}\sum_n P_n (2n+1)^2
\end{eqnarray}

\begin{itemize}
  \item {\it{Other vibrational modes}}
  \end{itemize}With N ions in the trap, assuming that the transversal potential is
strong enough to frozen the transversal degree of freedom,  only the
N longitudinal vibrational modes are relevant.  So far we have
assumed that only the collective center of mass motion is excited
and neglected other modes.  If we include the effect of other modes
into account, the fidelity is decreased. The main sources of
decoherence are :a) off resonant direct  couplings to other modes
and b) reduction of the coupling to the center of mass mode, $\chi$,
due to the vibration  of the other modes. However all these effects
are local and  the gap energetically suppresses them.

\begin{figure}[tbh]
\begin{center}
\epsfig{figure=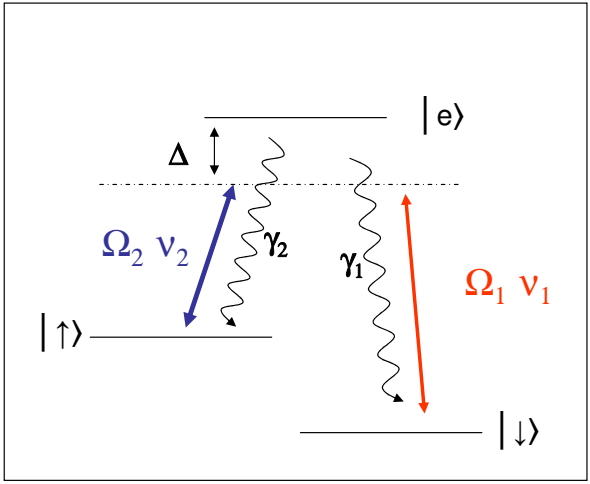, angle=270,scale=0.6}
\end{center}
\caption{Raman transition to a third level with atomic decay.
Here $\Delta$ is the
detuning of the laser  fields with frequencies $\omega_{e1,2}$ from the one photon resonance ,$\Delta=\omega_{e1}-\nu_1=\omega_{e2}-\nu_2$ ,
 and $\Omega_{1,2}$ are  laser  Rabi
frequencies.}\label{Ramman}
\end{figure}

\begin{itemize}
  \item {\it{Spontaneous emission}}
  \end{itemize}

  Additionally a more fundamental source of
decoherence arises from  spontaneous emission effects. In typical
ion trap experiments the $|\uparrow\rangle$ and $|\downarrow\rangle$
levels are coupled through  Raman transitions to a third excited
level $|e\rangle$ (see Fig.\ref{Ramman}). Assuming two photon
resonance conditions, that is
$\omega_{e1}-\omega_{e2}=\nu_1-\nu_2=\omega_0$ and
$\Delta=\omega_{e1}-\nu_1=\omega_{e2}-\nu_2$ where $\Delta$ is the
detuning of the fields from the one photon resonance and
$\omega_{e1,2}$ and $\Omega_{1,2}$ are  laser frequencies and Rabi
frequencies respectively, the Hamiltonian of the system in the
appropriate rotating frame can be written as
\begin{eqnarray}
\hat{H}_{s}&=& -\Delta \sum_i\hat{\sigma}^i_{ee}+\hat{H}_{Is}\\
\hat{H}_{Is}&=&\Omega_1\sum_i(\hat{\sigma}^i_{\downarrow e}
+\hat{\sigma}^i_{ e \downarrow})+
\Omega_2\sum_i(\hat{\sigma}^i_{\uparrow e} +\hat{\sigma}^i_{ e
\uparrow})
\end{eqnarray} where $\hat{\sigma}^i_{ee}=|e\rangle_i{}_i\langle e|$,
$\hat{\sigma}^i_{\downarrow e}=|\downarrow\rangle_i{}_i\langle e|$
and $\hat{\sigma}^i_{\uparrow e}=|\uparrow\rangle_i{}_i\langle e|$.

 The  decoherence processes due to spontaneous
emission can be described by means of Heisenberg-Langevin
equations\cite{Gardiner} given by:

\begin{eqnarray}
\dot{\hat{\sigma}}^j_{\uparrow \downarrow}&=&
=(\rm{i}\Omega_1+\hat{f}_{\downarrow
e}^{j\dagger})\hat{\sigma}^j_{\uparrow
e}-(\rm{i}\Omega_2-\hat{f}^j_{\uparrow
e})\hat{\sigma}^j_{e\downarrow}
\\ \dot{\hat{\sigma}}^j_{\uparrow
e}&=&-(\rm{i}\Delta+\Gamma_e/2)\hat{\sigma}^j_{\uparrow e} - (\rm{i}
\Omega_2- \hat{f}^j_{\uparrow e}) (\hat{\sigma}^j_{e
e}-\hat{\sigma}^j_{\uparrow \uparrow})\notag \\&&+(\rm{i} \Omega_1
-\hat{f}^j_{\downarrow
e})\hat{\sigma}^j_{\uparrow \downarrow} \\
\dot{\hat{\sigma}}^j_{e\downarrow }&=&
(\rm{i}\Delta-\Gamma_e/2)\hat{\sigma}^j_{e \downarrow}+(\rm{i}
\Omega_1 +\hat{f}_{\downarrow e}^{j\dagger})(\hat{\sigma}^j_{e
e}-\hat{\sigma}^j_{\downarrow \downarrow})\\&&-(\rm{i} \Omega_2
+\hat{f}_{\uparrow e}^{j\dagger})\hat{\sigma}^j_{ \uparrow
\downarrow}
\end{eqnarray}where $\Gamma_e=\gamma_1+\gamma_2$ with  $\gamma_1$ and $\gamma_2$ are decay rates form $|e\rangle$
to $|\downarrow \rangle$ and  $|\uparrow \rangle$  respectively and
the noise operators $\hat{f}$ have zero mean and are $\delta$
correlated\cite{Gardiner}:$ \overline{\langle \hat{f}^j_{ \downarrow
e}(t) \hat{f}^{k\dagger}_{ \downarrow e}(t')\rangle}=\gamma_1
\delta(t-t') \delta_{j,k}$ and $\overline{\langle
\hat{f}^j_{\uparrow e }(t) \hat{f}^{k \dagger}_{ \uparrow
e}(t')\rangle}=\gamma_2 \delta(t-t') \delta_{j,k}$.

 In the large
photon detuning limit $\Delta \gg \Omega_1 \Omega_2,\gamma_1,
\gamma_2$, one can adiabatically eliminate the operators
$\hat{\sigma}^j_{e\downarrow }$ and $\hat{\sigma}^j_{e\uparrow }$
and their hermite conjugates and then use the projected equations of
motion to solve for $\dot{\hat{\sigma}}^j_{\uparrow \downarrow}$.

\begin{eqnarray}
\dot{\hat{\sigma}}^j_{\uparrow \downarrow}&=& FH +i[\hat{H}_{\rm{
noise}},\hat{\sigma}^j_{\uparrow \downarrow}]\\
\hat{H}_{\rm{noise}}(t)&=&\frac{1}{2}\sum_j[h_{jz}^s(t)
\hat{\sigma}_j^z+h_{jx}^s(t) \hat{\sigma}_j^x+h_{jy}^s(t)
\hat{\sigma}_j^y
\end{eqnarray} where FH accounts for the the Hamiltonian part of the
dynamics and with $h_{jz}^s(t)=\frac{\Omega_1}{\rm{i}\Delta}(
\hat{f}^{j\dagger}_{\downarrow e}-\hat{f}^{j}_{\downarrow
e})-\frac{\Omega_2}{ \rm{i} \Delta}( \hat{f}^{j\dagger}_{\uparrow
e}-\hat{f}^{j}_{\uparrow e})$,
$h_{jx}^s(t)=-\frac{\Omega_2}{\rm{i}\Delta}(
\hat{f}^{j\dagger}_{\downarrow e}-\hat{f}^{j}_{\downarrow
e})-\frac{\Omega_1}{ \rm{i} \Delta}( \hat{f}^{j\dagger}_{\uparrow
e}-\hat{f}^{j}_{\uparrow e})$ and
$h^s_{jy}(t)=-\frac{\Omega_2}{\Delta}(
\hat{f}^{j\dagger}_{\downarrow e}+\hat{f}^{j}_{\downarrow
e})+\frac{\Omega_1}{ \Delta}( \hat{f}^{j\dagger}_{\uparrow
e}+\hat{f}^{j}_{\uparrow e})$.

From the previous expressions one can estimate the degradation of
the fidelity due to dephasing ( similar degradation of the fidelity
is caused by $x$ or $y$ type of noise). In the adiabatic limit, i.e.
$\Delta \gg \Omega_1 \Omega_2,\gamma_1, \gamma_2$, $h_{jz}^s$ are
independent stochastic white noise processes with zero mean and
autocorrelation function $\overline{h^{s}_{iz}(t)
h^{s}_{jz}(\tau)}=\gamma_{s}\delta_{ij}\delta(t-\tau)$ with
$\gamma_{sp}=\frac{(\gamma_1\Omega_1^2+\gamma_2\Omega_2^2)}{\Delta^2}$
and consequently  the gap can not protect against this broad band
noise. From Eq.(\ref{dec}) they will case  a degradation of the GHZ
fidelity of

\begin{eqnarray}
\mathcal{F}&=& 1-\gamma_{sp}N t_o
\end{eqnarray}

 In order to reduce the strong degradations due to this type of high-frequency
 decoherence processes one can increase
the Raman-detuning \cite{Ozeri} at the expense of slower evolution
which in turn will make the system more susceptible to other kind of
local noise (eg. magnetic field inhomogeneities). On the other hand
the latter can be suppressed by the MPM.

From this analysis we conclude that overhead of implementing
$\hat{J}^{(0)2}-\hat{J}_z^{(0)2}$ instead of $\hat{J}_z^{(0)2}$ is
mainly the additional  echo technique   required to remove the $n$
dependence of $H_{eff}$. Besides that, on average the same type of
non-ideal disturbances are found in both  Hamiltonians with the
advantage  of $\hat{J}^{(0)2}-\hat{J}_z^{(0)2}$  that  the gap
protects the system against those of them which are local in
character.

\section{Implementation in optical lattices}

\subsection{Engineering long-range interactions }

Up to now we have explored only  the generation of an MPM
  via isotropic long-range interactions.  In practice, however,
it is desirable to have a similar kind of protection  generated by
systems with short range interactions such as those provided by
 cold atoms in optical lattices. These systems offer the
possibility to dynamically change the Hamiltonian parameters at a
level unavailable in more traditional condensed matter systems. We
now show how an MPM can be created in lattice  systems and can be
 used  to robustly  generate
N-particle GHZ states.

We consider ultracold bosonic atoms with two relevant internal
states confined in a an
 optical lattice. We will
assume that the lattice is loaded with  one atom per site, and again
identify the two possible states of each site, with the effective
spin index $\sigma=\uparrow,\downarrow$ respectively. For
 deep periodic potential and low temperatures, the atoms
are confined to the lowest Bloch band and the low energy Hamiltonian
is  given by \cite{Jaksch}
\begin{eqnarray}
\hat{H}_{BH}&=& -\tau \sum_{\langle i,j\rangle \sigma}
\hat{a}_{\sigma,i}^{\dagger}\hat{a}_{\sigma,j}+\frac{1}{2}\sum_{j\sigma}
U_{\sigma \sigma}\hat{n}_{\sigma,j}(\hat{n}_{\sigma,j}-1) \notag
\\&&+
U_{\uparrow\downarrow}\hat{n}_{\sigma,j}\hat{n}_{\sigma,j}\label{EQNBH}
\end{eqnarray}

Here $\hat{a}_{\sigma,j}$ are bosonic annihilation operators of a
particle at site $j$ and state $\sigma$,
$\hat{n}_{\sigma,j}=\hat{a}_{\sigma,j}^{\dagger}\hat{a}_{\sigma,j}$,
and the sum $\langle i,j\rangle$ is over nearest neighbors. In
Eq.(\ref{EQNBH}) the  parameter $\tau$ is  the tunneling energy
between adjacent sites (which we assume spin independent i.e.  spin
independent lattices) and  and $U_{\sigma,\sigma'}$ are the
different  on-site interaction energies which depend on the
scattering length between the different species. Both
$U_{\sigma\sigma'}$ and $\tau$ are functions of the lattice depth.
We are interested in a unit filled lattice  in the regime $\tau \ll
U_{\sigma \sigma'}$ where the system is deep in the Mott insulating
phase \cite{Fisher,Greiner}. In this limit, to zero order in $\tau$
the ground state is multi-degenerate and corresponds to all possible
spin configuration with one atom per site. A finite $\tau$ breaks
the spin degeneracy. By including virtual particle-hole excitations
one can derive an effective Hamiltonian that describe the spin
dynamics within the one atom per site subspace\cite {Duan}:
\begin{equation}
\hat{H}_{lat}=\hat{H}_{H}+\hat{H}_{I}=-{\bar{\lambda}}\sum_{<i,j>,\alpha }\hat{%
\sigma}_{i}^{\alpha }\hat{\sigma}_{j}^{\alpha }-{\bar{\chi}}\sum_{<i,j>}\hat{%
\sigma}_{i}^{z}\hat{\sigma}_{j}^{z}.  \label{EQNBHH}
\end{equation}
Here the coefficients are  $\bar{\lambda}=\tau^{2}/U_{\uparrow \downarrow }$ and $\bar{\chi}%
=\tau^{2}(U_{\uparrow \uparrow }^{-1}+U_{\downarrow \downarrow
}^{-1} -2U_{\uparrow \downarrow }^{-1})$. For simplicity we will now
restrict the analysis to one dimensional systems and assume periodic
boundary conditions.

 $\hat{H}_{H}$ is spherically symmetric and in terms of collective spin operators it can be written as
\begin{equation}
\hat{H}_{H}=-\frac{4\bar{\lambda}}{N}  \hat{J}^{(0)2}-\frac{4\bar{\lambda}}{N}\sum_{k=1\dots N-1, \alpha }{\hat{
J}_{\alpha}^{(k) }}{\hat{
J}_{\alpha}^{(-k) }}  \cos\left(\frac{2 \pi k }{N}\right)\label{bas}
\end{equation}
 All the  $N+1$  fully
symmetric states with $J=N/2$ are degenerate and span  the ground state of
 $\hat{H}_{H}$. $\hat{H}_{I}$ is not spherically symmetric but we can also write  it in terms of
collective operators as
\begin{equation}
\hat{H}_{I}=-\frac{4\bar{\chi}}{N}  \hat{J}^{(0)2}_{z}-\frac{4\bar{\chi}}{N}\sum_{k=1\dots N-1}\hat{
J}_{z}^{(k) } \hat{
J}_{z}^{(-k) } \cos\left(\frac{2 \pi k }{N}\right).\label{bas}
\end{equation}
If the condition $\bar{\chi}\ll \bar{\lambda}$ is satisfied, which
can be engineered in this atomic systems by means of a Feshbach
resonance, the effect of the Ising term
 can be studied   by means of perturbation theory. Assuming that at $t=0$ the initial
state is prepared within the $J=N/2$ manifold, a  perturbative analysis predicts
 that for times $t$ such that $\bar{\chi}t<\bar{\lambda}/\bar{\chi}$,
 $\hat{H}_{H}$ confines the dynamics to the ground state manifold  and
transitions outside it   can be neglected. As a consequence, only
the projection of  $\hat{H}_{I}$  on it, which corresponds   to
 $\mathcal {P}\hat{H}_{I}= {\chi }_{e} \hat{J}^{(0)2}_{z}-\frac{\bar{\lambda} N}{ N-1}$ with
${\chi }_{e}\equiv \frac{4\bar{\chi}}{N-1}$, is effective  and $H_I$
acts as a long range Hamiltonian. Here we used the relation
$\mathcal{P}_{k \neq 0}[\hat{ J}_{z}^{(k)} \hat{ J}_{z}^{(-k)
}]=-\frac{\hat{ J}_z{(0)^2}}{N-1} +\frac{N^2}{4 (N-1)}$, with
$\mathcal{P}$ the projection into the $J=N/2$ subspace.
 The non zero projection of the latter term  comes from the fact that the operators $\hat{
J}_{z}^{(k)} \hat{ J}_{z}^{(-k) }$ and ${\hat{ J}_{z}} ^{(0)2}$ are
not independent as they satisfy the constrain $\sum_{k=0}^{N-1}
\hat{ J}_{z}^{(k) } \hat{ J}_{z}^{(-k) }=N^2/4$.

\begin{figure}
\includegraphics[width=3.3 in,height=2.5 in]{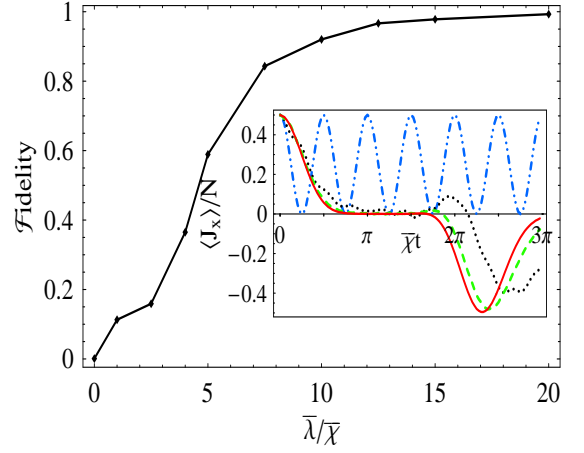}
 \caption{(color online)  Fidelity to generate a GHZ state vs $\bar{\lambda}/\bar{\chi}$. In
the inset we show $\langle{\hat{J}^{(0)}}_x(t)\rangle $. The blue
dot-dashed, dotted black,  dashed green and solid red correspond to
$\bar{\lambda} = 0, 5, 10, 20$ respectively. The plots are obtained
by  numerical evolution of Eq. (\ref{EQNBHH}) for  $N=10$.
}\label{latfi1}
\end{figure}

In Fig. \ref{latfi1} we contrast the dynamical evolution of a system
in the presence and absence of $\hat{H}_H$ assuming  at time $t=0$
all the spins are polarized in the $x$ direction. If  only the Ising
term is present, $\bar{\lambda}=0$, it induces local phase
fluctuations that leads to fast oscillations in
$\langle{\hat{J}^{(0)}}_x\rangle=N/2 \cos^2[2 \bar{\chi} t]$. On the
other hand, as the ratio  $\bar{\lambda} / \bar{\chi}$ increases,
the isotropic interaction inhibits the fast oscillatory dynamics
 and instead
$\langle{\hat{J}^{(0)}}_x\rangle$ exhibits slow collapses and revivals. For $\bar{\lambda} \gg {\chi}$ the dynamics
exactly resembles the one induced by $\hat{H}_z$ and  at ${\chi }_{e} t=\pi/2$ the initial coherent state is squeezed into
a $GHZ$ state.

\subsection{MPM in lattice systems}

$\hat{H}_{H}$ also provides protection against phase decoherence. However, $\hat{H}_{H}$
is not as effective as  $\hat{H}_{prot}$ because
 the energy gap between the MPM and the excited states
 of $\hat{H}_H$  vanishes in the thermodynamic limit  as $E_g \to \bar{\lambda}/N^2$ . This is a
  drawback  of  the short
range Hamiltonian for the purpose of  fully protecting the ground
states from long wave length excitations. Note however that one
dimensional systems are the worst scenario as for higher dimensions
the gap vanishes  as $N^{-2/d}$ with $d$ the dimensionality of the
system. Nevertheless, the many body interactions  can
 still  eliminate short-wavelength excitations   since in the large $N$ limit they remain separated by a finite
energy  gap, $8 \bar{\lambda}$.

We quantify the effectiveness of the MPM to protect the system
against $\hat{H}_{env}$  by using time dependent perturbation
theory. For this analysis we restrict to the  limit $\bar{\lambda}
\gg \bar{\chi}$ where the Ising term can be treated as an effective
${\chi }_{e} \hat{J}^{(0)2}_{z}-\frac{\bar{\lambda} N}{N-1}$
Hamiltonian. In this limit  a convenient basis to study the quantum
dynamics is  the collective spin basis. Assuming that at $t=0$ the
system lies within the $J=N/2$ manifold, the evolution of the matrix
elements  $ \rho_{M \tilde{M}}\equiv {}_z\langle
N/2,\tilde{M}|\hat{\rho}| N/2,M \rangle_z$ can be written as
\begin{equation}
 \rho_{M,\tilde{M}}(t)= \rho_{M,\tilde{M}}(0) e^{i t {\chi }_{e} (M^2-\tilde{M}^2)}
 e^{i(\theta_M-\theta_{\tilde{M}}}) e^{-\frac{1}{2}(\gamma^M_{lat}-\gamma^{\tilde{M}}_{lat}}) \label{denspl}
\end{equation}
where the random  phase, given by Eq.(\ref{thet}), characterizes the
 dynamics induced by the noise within the MPM
and $\gamma_{lat} ^{M}(t)=\sum_{J\neq{N/2},\beta }|\int_{0}^{t}d\tau
\mathcal{M}_{J,\beta }^{M}e^{i\tau \omega_{J,\beta}^{lat}}|^{2}$,
takes into account the depletion of the $J=N/2$ levels due to
 transition  matrix elements with states outside the symmetric manifold:
 $\mathcal{M}_{J,\beta }^{M}={}_z\langle
\frac{N}{2},M|\hat{H}_{env}|J,M,\beta \rangle {}_z$.
$\omega^{lat}_{J,\beta}$ are the respective energy splittings. Up to
this point the expressions are structurally identical to the ones
obtained for long range interactions. The difference appears in the
evaluation of $\gamma_{lat}$.  In contrast  to $\hat{H}_{prot}$, not
only  the excitation frequencies
 $\omega^{lat}_{J,\beta}$   are not  degenerated but also they become smaller as $N$ is increased.
 As a consequence  Eq. (\ref{gampr})
 is replaced by the following equation for the lattice system
\begin{eqnarray}
\overline{ \gamma_{lat} ^{M}(t)}\approx \frac {N^2-4 M^2}{ N(N-1)}f
\sum_{k=1}^{N-1}\int_0^{\omega_c} d\omega
\left(\frac{\sin(t(\omega-\Delta E^k)/2)}{\omega-\Delta
E^k}\right)^2 \label{lardec}
\end{eqnarray}
 with $E^k$ the excitation energies  of the states
that belong to the  $J=N/2-1$ manifold given by $\Delta E^{k}=8\bar{\lambda}\sin ^{2}(\pi k/N)$, with $%
k=1,\dots ,N-1$ \cite{bethe}. From Eq.(\ref{lardec}) we can estimate
the degradation of the fidelity due to phase decoherence as:

\begin{figure}[tbh]
\begin{center}
\leavevmode {\includegraphics[width=3. in]{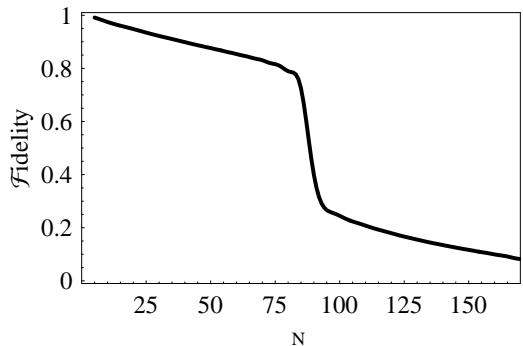}}
\end{center}
\caption{  In the presence of phase decoherence the fidelity of the
GHZ state preparation in the lattice is degraded as  $N$ grows
because in the lattice the gap decreases and the generation time
increases with $N$. In this plot we assumed the limit $\bar{\chi}\ll
\bar{\lambda}$ where Eq. (\ref {fidla}) holds and used a
 system with
$\omega_c= \bar{\chi}$, $\bar{\lambda}=100 \bar{\chi}$ and
$\Gamma=0.01 \bar{\chi}$.  At $N=90$, $\Delta E_g=\omega_c$ and it
explains  the drop of $\mathcal{F}$ for $N>90$.} \label{lastf}
\end{figure}

\begin{equation}
\mathcal{F}(t_o)\gtrsim e^{\overline{ \gamma_{lat}
^{0}(t)}}\frac{1}{\sqrt{1+\Gamma(t_o)}}\label{fidla}.
\end{equation} In Fig.\ref{lastf} we plot $\mathcal{F} $ calculated from Eq.(\ref{fidla})as a
function of $N$.  In the lattice the fidelity  is degraded as $N$
grows because the gap decreases and the generation time increases
with $N$. Moreover,  an abrupt drop of the fidelity occurs at the
value of $N$ at which $ E_g =\omega_c$.

\section{Noise and decoherence in lattice systems }

In the previous section we used the effective Hamiltonian given by
Eq. (\ref{EQNBHH}) to study the GHZ generation in lattice systems.
Here we perform a more detailed analysis of its validity  and
estimate the effect of deviations from the ideal situations in an
actual experiment.  For this analysis we restrict to the  limit
$\bar{\lambda} \gg \bar{\chi}$ where the Ising term can be treated
as an effective ${\chi }_{e}
\hat{J}^{(0)2}_{z}-\frac{\bar{\lambda}N}{N-1}$ Hamiltonian.

\subsection{Particle-hole excitations}

Deriving Eq. (\ref{EQNBHH}) from the Bose-Hubbard Hamiltonian we
only included virtual-particle hole excitation. However, during the
time evolution real transitions  from singly to doubly occupied
states can take place and they  degrade the fidelity.

To account for these effects, we write  the manybody wave function
as $|\Psi(t)\rangle=\sum_{n} C_n |\psi_n\rangle+
\sum_{m}B_m|\phi_m\rangle$, where $|\psi_n\rangle$ span the Hilbert
space with one atom per site and $|\phi_m\rangle$  span the subspace
with one  particle and one hole adjacent to each other and $N-2$
singly occupied sites. The latter  are the states that directly
coupled to $|\psi_n\rangle$ through tunneling. Solving the time
dependent Schr\"{o}dinger equation  from the Bose Hubbard
Hamiltonia, using the assumption that $U_{\sigma,\sigma'}\approx U$
and that at time $t=0$ no doubly occupied states are populated one
obtains

\begin{eqnarray}
i \dot{C}_n=\sum_{k}\langle
\psi_n|\hat{H}_{latt}|\psi_k\rangle(1-e^{-iUt})C_k \label{eqph}
\end{eqnarray}Here we also assumed that  $\{C_n\}$ change at a rate
much smaller than $U_{\sigma,\sigma'}$ and treated them as constants
during the time integration. Eq.(\ref{eqph}) yields  the following
lost of fidelity due to real  particle hole excitations:

\begin{equation}
\mathcal{F}(t_o)\approx 1- \frac{4 \bar{\chi}}{U}\sin^2(U
t_o/2)\label{fidph}.
\end{equation}remembering that $\chi_e t_o=\pi/2$. As long as $\bar{\chi}/U
\ll 1$, we conclude that particle-hole excitations do not
significantly affect the GHZ generation.

\subsection{Magnetic confinement}

In Eq. (\ref{EQNBH}) we assumed a translationally invariant system.
However in most of the experiments an additional quadratic magnetic
confinement is used to collect the atoms. Actually is due to this
quadratic potential that a unit filled Mott insulator has  been
experimentally realized. In its absence it would be difficult to
create
 a unit filled Mott insulator as  in an homogeneous system it only takes
place   when the number of atoms is exactly equal to the number of
lattice sites. A drawback of the magnetic confinement is that it
generates always superfluid regions at the edge of the cloud, so
only  a fraction of the total trapped atoms located at the trap
center has to be selected as the quantum register. Assuming we work
on this unit filled Mott Insulator subspace, here we quantify the
effect of the magnetic potential in the GHZ generation in  the
$\bar{\lambda}/\bar{\chi} \gg 1$ limit.

The magnetic confinement is accounted for by adding a term $W
\sum_{j,\sigma} j^2 \hat{n}_{\sigma,j}$ in the Bose Hubbard
Hamiltonian. $ W=1/2 m \omega_T^2 a_L^2$ with  $m$ the atom mass,
$\omega_T$ the frequency of the external trapping potential and
$a_L$ the lattice spacing. This term modifies the global coupling
constants $\bar{\lambda}$ and $\bar{\chi}$ when the effective
Hamiltonian is derived  and make them site dependent, $\bar{\lambda}
\to \bar{\lambda}_i^W\equiv\tau^{2}/\widetilde{U}_{i,\uparrow
\downarrow }$ and  $\bar{\chi} \to
\bar{\chi}_i\equiv\tau^{2}(\widetilde{U}_{i,\uparrow \uparrow
}^{-1}+\widetilde{U}_{i,\downarrow \downarrow }^{-1}
-2\widetilde{U}_{i,\uparrow \downarrow }^{-1})$. Here
$\widetilde{U}_{i,\sigma\sigma' }=U_{\sigma\sigma'
}/(U^2_{\sigma\sigma' }-W^2(2i+1)^2)$. Assuming that the gradient of
the external potential is weak compared to the on site interaction
energy, as is in general the case for current experiments, the
effective Hamiltonian in the presence of the magnetic trap becomes

\begin{eqnarray}
\hat{H}_{lat}^W&=&\hat{H}_{lat}+\hat{H}_1^T\\
\hat{H}_1^T&=&-\sum_{{<i,j>}}{T_i}\vec{\sigma}_i\cdot\vec{\sigma}_{j}\\
{T_i}&=&-\frac{\tau^2 W^2(2i+1)^2}{U^3}
 \label{spin2}
\end{eqnarray}

The  corrections on the fidelity of the GHZ state
 introduced by  $\hat{H}_1^T$ can be estimated by calculating the effective projection of it on the MPM:   $\mathcal {P}
\hat{H}_1^T$ . As the latter is just proportional to the identity
matrix $I$, $\mathcal {P} \hat{H}_1^T=\sum_i T_i I $, it effects is
just a global phase and it does not cause any main degradation of
the fidelity. Similarly any other  perturbation induced by local
fluctuations in the magnetic field or the lasers used to generate
the lattice become irrelevant thanks to the MPM.

From this analysis we conclude that except from spontaneous emission
or heating mechanisms the MPM effectively protects lattice systems
against common non-ideal situations encountered during their
experimental realization.  On the other hand  lattice-based GHZ
state generation faces   the scalability problem due to the fact
that the gap decreases and the generation time increases with
increasing $N$.

\section{Conclusions}

We have in this paper evaluated the possibility for a robust
preparation of multi-particle GHZ entangled states of trapped ions
or  cold atoms in optical lattice by generating a decoherence free
multi-level manifold corresponding  to the ground levels of properly
designed Hamiltonians.  The MPM is isolated from the rest of the
Hilbert space by an energy gap which energetically suppresses  any
local decoherence processes. We have presented analytical estimates
for the fidelity of the GHZ preparation.

 In trapped ions we
demonstrated that the fidelity can be significantly  better than the
one achievable without any gap protection and therefore that our
scheme is in the position to improve the spectroscopy resolution in
current Ramsey spectroscopy experiments.

We also showed that cold atoms in optical lattices interacting via
short range interactions can be utilized  to engineer  long range
interactions which in turn can be used for generating many-body
entanglement. We calculated the effects of non-ideal conditions and
concluded that the main restriction in these systems is the
scalability as the MPM protection degrades with increasing $N$.

The scalability certainly limits the use of lattice systems for
massive  entanglement generation, however it  is not a problem for
recent quasi-one dimensional experiments \cite{ Paredes} where an
array of 1D tubes with an average of 18 atoms per tube has been
realized.  In such systems therefore it should be possible to create
few-particle collective entangled states using our scheme
 and to perform  proof-of-principle experiments demonstrating
  the improvement of spectroscopic sensitivity.

We emphasize  that, even though we have limited the discussion to
ensembles of spin $S= 1/2$ particles, the  MPM ideas can be
straightforwardly
 generalized to systems composed of  higher spin atoms.
Besides entanglement generation,  the MPM  might have also important
applications for  the implementation of good storage memories using
 for example nuclear spin ensembles in solid state \cite{Johnson} or
 photons\cite{Fleischhauer}.

\section{ Acknowledgements} This work was supported by ITAMP, NSF
(Career Program), AFOSR, ONR MURI and the David and Lucille Packard
Foundation.

\bibliographystyle{plain}

\end{document}